\documentclass[journal]{vgtc}                %
\ifpdf%
  \pdfoutput=1\relax                   %
  \usepackage{graphicx}                %
  \DeclareGraphicsExtensions{.pdf,.png,.jpg,.jpeg} %
\else%
  \usepackage{graphicx}                %
  \DeclareGraphicsExtensions{.eps}     %
\fi%

\graphicspath{{figures/}{pictures/}{images/}{./}} %

\usepackage{microtype}                 %
\PassOptionsToPackage{warn}{textcomp}  %
\usepackage{textcomp}                  %
\usepackage{mathptmx}                  %
\usepackage{times}                     %
\usepackage{cite}                      %
\usepackage{tabu}                      %
\usepackage{booktabs}                  %
\usepackage{enumitem}
\usepackage{wrapfig}
\usepackage[hidelinks]{hyperref}
\usepackage{xcolor}
\usepackage{graphbox}
\usepackage{dblfloatfix}

\onlineid{0}

\vgtccategory{Research}
\vgtcpapertype{please specify}

\title{\tool{}: Learning Convolutional Neural Networks with Interactive Visualization}

\author{Zijie J. Wang, Robert Turko, Omar Shaikh, Haekyu Park, Nilaksh Das, \\Fred Hohman, Minsuk Kahng, and Duen Horng (Polo) Chau}
\authorfooter{
\item
 Zijie J. Wang, Robert Turko, Omar Shaikh, Haekyu Park, Nilaksh Das, Fred Hohman, and Duen Horng Chau are with Georgia Tech. E-mail: \{jayw$\mid$rturko3$\vert$oshaikh$\mid$haekyu$\mid$nilakshdas$\mid$fredhohman$\mid$polo\}@gatech.edu.
\item
 Minsuk Kahng is with Oregon State University. E-mail: minsuk.kahng@oregonstate.edu.
}

\shortauthortitle{Zijie J. Wang \MakeLowercase{\textit{et al.}}: CNN Explainer}

\newcommand{\tool}[0]{\textsc{CNN Explainer}}
\newcommand{\quotes}[1]{``\textit{#1}''}

\newcommand{\overview}[0]{Overview}
\newcommand{\elastic}[0]{Elastic Explanation View}
\newcommand{\elasticConv}[0]{Convolutional Elastic Explanation View}
\newcommand{\elasticFlatten}[0]{Flatten Elastic Explanation View}
\newcommand{\formula}[0]{Interactive Formula View}
\newcommand{\formulaConv}[0]{Convolutional Interactive Formula View}

\newcommand{\formulaSoft}[0]{Softmax Interactive Formula View}
\newcommand{\widgetParam}[0]{Hyperparameter Widget}

\definecolor{orange}{RGB}{255,119,0}
\definecolor{red}{RGB}{220,0,0}
\definecolor{agreen}{RGB}{74, 198, 148}
\definecolor{purple}{RGB}{158, 62, 177}
\definecolor{darkpurple}{RGB}{170, 70, 210}
\definecolor{aqua}{RGB}{87, 180, 181}
\definecolor{lightblue}{RGB}{72, 123, 232}
\definecolor{hotpink}{RGB}{255, 83, 115}
\definecolor{teal}{RGB}{90, 200, 250}
\definecolor{linkColor}{RGB}{6,125,233}
\definecolor{RdColorscale}{RGB}{213, 96, 80}
\definecolor{BuColorscale}{RGB}{75, 148, 196}
\definecolor{BrColorscale}{RGB}{188, 132, 53}
\definecolor{BGColorscale}{RGB}{57, 152, 143}
\definecolor{LOrangesColorscale}{RGB}{255, 195, 133}
\definecolor{ROrangesColorscale}{RGB}{196, 65, 3}
\definecolor{layerGray}{RGB}{80, 80, 80}

\newcommand{\layer}[1]{\textcolor{layerGray}{\textbf{\texttt{\detokenize{#1}}}}}

\abstract{
Deep learning's great success motivates many practitioners and students to learn about this exciting technology.
However, it is often challenging for beginners to take their first step due to the complexity of understanding and applying deep learning.
We present \tool{}, 
an interactive visualization tool 
designed for non-experts to learn and examine convolutional neural networks (CNNs), a foundational deep learning model architecture.
Our tool addresses key challenges that novices face while learning about CNNs, which we identify from interviews with instructors and a survey with past students.
\tool{} tightly integrates 
a model overview that summarizes a CNN's structure, 
and on-demand, dynamic visual explanation views that help users understand the underlying components of CNNs.
Through smooth transitions across levels of abstraction, 
our tool enables users to inspect the interplay between low-level mathematical operations and high-level model structures.
A qualitative user study shows that \tool{} helps users more easily understand the inner workings of CNNs, and is engaging and enjoyable to use.
We also derive design lessons from our study.
Developed using modern web technologies, \tool{} runs locally in users' web browsers without the need for installation or specialized hardware, broadening the public's education access to modern deep learning techniques.
}


\CCScatlist{ %
 \CCScat{K.6.1}{Management of Computing and Information Systems}%
{Project and People Management}{Life Cycle};
 \CCScat{K.7.m}{The Computing Profession}{Miscellaneous}{Ethics}
}

\teaser{
  \centering
  \includegraphics[width=\linewidth]{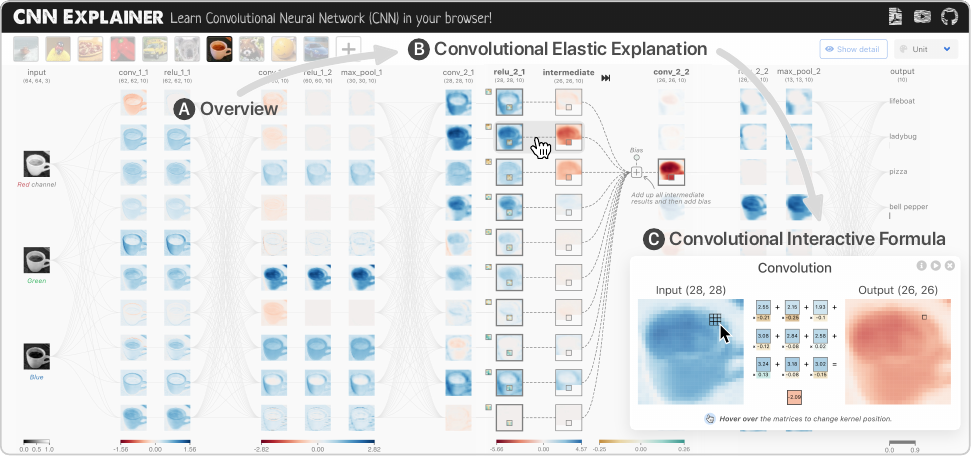}
  \setlength{\abovecaptionskip}{-5pt}
  \caption{
  With \tool{}, learners 
  can visually 
  examine how \textit{Convolutional Neural Networks} (CNNs) transform input images into classification predictions (e.g., predicting \textit{espresso} for an image of a coffee cup), and interactively learn about their underlying mathematical operations.
  In this example, 
  a learner uses \tool{} to understand 
  how convolutional layers work through
  three tightly integrated views, each explaining the convolutional process in increasing levels of detail.
  \textbf{(A)} The \textit{\overview{}} visualizes a CNN architecture where each neuron is encoded as a square with a heatmap representing the neuron's output.
  \textbf{(B)} Clicking a neuron reveals how its activations are computed by the previous layer's neurons, displaying the often-overlooked intermediate computation through animations of sliding kernels.
  \textbf{(C)} 
  \textit{\formulaConv{}} for inspecting
  underlying mathematics of the dot-product operation core to convolution.
  For clarity, some annotations are removed and views are re-positioned.
  }
  \label{fig:crown}
}

\vgtcinsertpkg

\begin{document}

\firstsection{Introduction}

\maketitle

Deep learning now enables many of our everyday technologies.
Its continued success 
and 
potential 
application
in diverse domains has attracted immense interest from students and practitioners who wish to learn and apply this technology.
However, many beginners find it challenging to take the first step in studying and understanding deep learning concepts.
For example, convolutional neural networks (CNNs), a foundational deep learning model architecture, is often one of the first and most widely used models that students learn.
CNNs are often used in image classification, achieving state-of-the-art performance \cite{lecunDeepLearning2015}.
However, through interviews with deep learning instructors and a survey of past students, we found that even for this ``introductory'' model, it can be challenging for beginners to understand how inputs (e.g., image data) are transformed into class predictions.
This steep learning curve stems from CNN's complexity, which typically leverages many computational layers to reach a final decision.
Within a CNN, there are many types of network layers (e.g., fully-connected, convolutional, activation), each with a different structure and underlying mathematical operations.
Thus, a student needs to develop a mental model of not only how each layer operates, but also how to choose different layers that work together to transform data.
Therefore, a key challenge in learning about CNNs is the intricate interplay between \textit{low-level mathematical operations} and \textit{high-level integration} of such operations within the network.

\textbf{Key challenges in designing learning tools for CNNs.}
There is a growing body of research that uses interactive visualization to explain the complex mechanisms of modern machine learning algorithms, such as TensorFlow Playground \cite{smilkovDirectManipulationVisualizationDeep2017} and GAN Lab \cite{kahngGANLabUnderstanding2019}, which help students learn about dense neural networks and generative adversarial networks (GANs) respectively.
Regarding CNNs, some existing visualization tools focus on demonstrating the high-level model structure and connections between layers (e.g., Harley's Node-Link Visualization \cite{harleyInteractiveNodeLinkVisualization2015a}), while others focus on explaining the low-level mathematical operations (e.g., Karpathy's interactive CNN demo \cite{karpathyConvNetJSMNISTDemo2016}).
There is no visual learning tool that explains and connects CNN concepts from both levels of abstraction.
This interplay between global model structure and local layer operations has been identified as one of the main obstacles to learning deep learning models, as discussed in \cite{smilkovDirectManipulationVisualizationDeep2017} and corroborated from our interviews with instructors and student survey.
\tool{} aims to bridge this critical gap.

\begin{figure}[tb]
    \centering
    \includegraphics[width=\columnwidth]{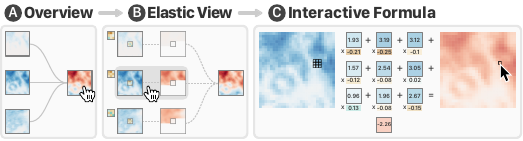}
    \caption{
    In \tool{}, tightly integrated views with different levels of abstractions work together to help users more easily learn about the intricate interplay between a CNN's high-level structure and low-level mathematical operations.
    \textbf{(A)} the \textit{Overview} summarizes connections of all neurons;
    \textbf{(B)} the \textit{Elastic View} animates the intermediate convolutional computation of the user-selected neuron in the \textit{Overview};
    and \textbf{(C)} \textit{Interactive Formula} interactively demonstrates the detailed calculation on the selected input in the \textit{Elastic View}.
    }
    \label{fig:intro-figure}
\end{figure}

\textbf{Contributions.} In this work, we contribute:

\begin{itemize}[topsep=0mm, itemsep=0mm, parsep=1mm, leftmargin=6mm]

\item \textbf{\tool{}, an interactive visualization tool designed for non-experts} to learn about both CNN's high-level model structure and low-level mathematical operations,
addressing learners' key challenge in connecting unfamiliar layer mechanisms with complex model structures. 
Our tool advances over prior work \cite{harleyInteractiveNodeLinkVisualization2015a, karpathyConvNetJSMNISTDemo2016}, overcoming unique design challenges identified from a literature review, instructor interviews and a survey with past students (\autoref{sec:design-challenge}).

\item \textbf{Novel interactive system design} of \tool{} (\autoref{fig:crown}), which
adapts familiar techniques such as \textit{overview + detail} and \textit{animation} to simultaneously summarize intricate model structure, while providing context for users to inspect detailed mathematical operations.
\tool{}'s visualization techniques work together through fluid transitions between different abstraction levels (\autoref{fig:intro-figure}), helping users gain a more comprehensive understanding of complex concepts within CNNs (\autoref{sec:interface}).
\item \textbf{Design lessons distilled from user studies} on an interactive visualization tool for machine learning education.
While visual and interactive approaches have been gaining popularity in explaining machine learning concepts to non-experts, little work has been done to evaluate such tools \cite{kahngHowDoesVisualization2019, nortonAdversarialPlaygroundVisualizationSuite2017a}.
We interviewed four instructors who have taught CNNs and conducted a survey with 19 students who have previously learned about CNNs to identify the needs and challenges for a deep learning educational tool (\autoref{sec:design-challenge}).
In addition, we conducted an observational study with 16 students to evaluate the usability of \tool{}, and investigated how our tool could help students better understand CNN concepts (\autoref{sec:user-study}).
Based on these studies, we discuss the advantages and limitations of interactive visual educational tools for machine learning.

\item \textbf{An open-source, web-based implementation} that broadens the public's education access to modern deep learning techniques without the need for advanced computational resources. 
Deploying deep learning models conventionally requires significant computing resources, e.g., servers with powerful hardware.
In addition, even with a dedicated backend server, it is challenging to support a large number of concurrent users.
Instead, \tool{} is developed using modern web technologies, where all results are directly and efficiently computed in users' web browsers (\autoref{sec:implementation}).
Therefore, anyone can access \tool{} using their web browser without the need for installation or a specialized backend.
Our code is open-sourced\footnote{Code:  \textcolor{linkColor}{\url{https://github.com/poloclub/cnn-explainer}}} and \tool{} is available at the following public demo link: \textcolor{linkColor}{\url{https://poloclub.github.io/cnn-explainer}}.

\end{itemize}

\textbf{Broadening impact of visualization for AI.}
In recent years, many visualization systems have been developed for deep learning, but very few are designed for non-experts \cite{harleyInteractiveNodeLinkVisualization2015a, kahngGANLabUnderstanding2019, olahNeuralNetworksManifolds2014, smilkovDirectManipulationVisualizationDeep2017}, as surveyed in \cite{hohmanVisualAnalyticsDeep2019}.
\tool{} joins visualization research that introduces beginners to modern machine learning concepts.
Applying visualization techniques to explain the inner workings of complex models has great potential.
We hope our work will inspire further research and development of visual learning tools that help democratize and lower the barrier to understanding and applying artificial intelligent technologies.

\section{Background for Convolutional Neural Networks}
This section provides a high-level overview of convolutional neural networks (CNNs) in the context of image classification, which will help ground our work throughout this paper.

Image classification has a long history in the machine learning research community.
The objective of supervised image classification is to map an input image, $X$, to an output class, $Y$. 
For example, given a cat image, a sophisticated image classifier would output a class label of ``cat''.
CNNs have demonstrated state-of-the-art performance on this task, in part because of their multiple layers of computation that aim to learn a better representation of image data.

CNNs are composed of several different layers (e.g., convolutional layers, downsampling layers, and activation layers)---each layer performs some predetermined function on its input data.
Convolutional layers ``extract features'' to be used for image classification, with early convolutional layers in the network extracting low-level features (e.g., edges) and later layers extracting more-complex semantic features (e.g., car headlights).  
Through a process called backpropagation, a CNN learns kernel weights and biases from a collection of input images.
These values also known as parameters, which summarize important features within the images, regardless of their location.
These kernel weights slide across an input image performing an element-wise dot-product, yielding intermediate results that are later summed together with the learned bias value.
Then, each neuron gets an output based on the input image.
These outputs are also called activation maps.
To decrease the number of parameters and help avoid overfitting, CNNs downsample inputs using another type of layer called pooling.
Activation functions are used in a CNN to introduce non-linearity, which allows the model to learn more complex patterns in data.
For example, a Rectified Linear Unit (ReLU) is defined as $\max\left(0, x\right)$, which outputs the positive part of its argument.
These functions are also often used prior to the output layer to normalize classification scores, for example, the activation function called Softmax performs a normalization on unscaled scalar values, known as logits, to yield output class scores that sum to one.
To summarize, compared to classic image classification models that can be over-parameterized and fail to take advantage of inherent properties in image data, CNNs create spatially-aware representations through multiple stacked layers of computation. 
\section{Related Work}

\subsection{Visualization for Deep Learning Education}

Researchers and practitioners have been developing visualization systems that aim to help beginners learn about deep learning concepts.
Teachable Machine \cite{carneyTeachableMachineApproachable2020} teaches the basic concept of machine learning classification, such as overfitting and underfitting, by allowing users to train a deep neural network classifier with data collected from their own webcam or microphone.
The Deep Visualization Toolbox \cite{yosinskiUnderstandingNeuralNetworks2015} also uses live webcam images to interactively help users to understand what each neuron has learned.
These deep learning educational tools feature direct model manipulation as core to their experience.
For example, users learn about CNNs, dense neural networks, and GANs through experimenting with model training in ConvNetJS MNIST demo \cite{karpathyConvNetJSMNISTDemo2016}, TensorFlow Playground \cite{smilkovDirectManipulationVisualizationDeep2017}, and GAN Lab \cite{kahngGANLabUnderstanding2019}, respectively.
Beyond 2D visualizations, Node-Link Visualization \cite{harleyInteractiveNodeLinkVisualization2015a} and TensorSpace\cite{TensorSpaceJsNeural2018} demonstrate deep learning models in 3D space.
Inspired by Chris Olah's interactive blog posts \cite{olahNeuralNetworksManifolds2014}, interactive articles explaining deep learning models with interactive visualization are gaining popularity as an alternative medium for education \cite{carterUsingArtificialIntelligence2017,madsenVisualizingMemorizationRNNs2019}.

Most existing educational resources focus on explaining either the high-level model structures and training process or the low-level mathematics, but not both.
However, we found that one key challenge for beginners learning about deep learning models is the difficulty connecting unfamiliar layer mechanisms with complex model structures (discussed in Sect. 4).
For example, TensorFlow Playground \cite{smilkovDirectManipulationVisualizationDeep2017}, one of the few yet popular deep learning educational tools, focuses on helping users develop intuition about the effects of different \textit{dense neural network} architectures, but does not explain the underlying mathematical operations.
TensorFlow Playground also operates on synthetic 2D data, which can be challenging for users to transfer newly learned concepts to more realistic data and models.
In comparison, our work explains both model structure and mathematics of \textit{CNNs}, a more complex architecture, with real image data.

\subsection{Algorithm Visualization}

Before deep learning started to attract interest from students and practitioners, visualization researchers have been studying how to design algorithm visualizations (AV) to help people learn about dynamic behavior of various algorithms \cite{hundhausenMetaStudyAlgorithmVisualization2002, shafferAlgorithmVisualizationState2010, brownAlgorithmAnimation1988}.
These tools often graphically represent data structures and algorithms using interactive visualization and animations \cite{gallesDataStructureVisualizations2006, guoOnlinePythonTutor2013, brownAlgorithmAnimation1988}.
While researchers have found mixed results on AV's effectiveness in computer science education \cite{fouhRoleVisualizationComputer2012, grissomAlgorithmVisualizationCS2003, byrneEvaluatingAnimationsStudent1999},
growing evidence has shown that student engagement is the key factor for successfully applying AV in education settings \cite{napsExploringRoleVisualization2002, hundhausenUsingVisualizationsLearn2000}.
Naps, et al. defined a taxonomy of six levels of engagement\footnote{Six engagement categories: \textit{No Viewing}, \textit{Viewing}, \textit{Responding}, \textit{Changing}, \textit{Constructing}, \textit{Presenting}.} at which learners can interact with AVs\cite{napsExploringRoleVisualization2002}, and studies have shown higher engagement level leads to better learning outcomes \cite{hansenDesigningEducationallyEffective2002, fouhRoleVisualizationComputer2012, schweitzerInteractiveVisualizationActive2007, kehoeRethinkingEvaluationAlgorithm2001}.

Deep learning models can be viewed as specialized algorithms comprised of complex and stochastic interactions between multiple different computational layers.
However, there has been little work in designing and evaluating visual educational tools for deep learning in the context of AV.
\tool{}'s design draws inspiration from the guidelines proposed in AV literature (discussed in \autoref{sec:design-goal}); our user study results also corroborate some of the key findings in prior AV research (discussed in \autoref{sec:study-result}).
Our work advances AV's landscape in covering modern and pervasive machine learning algorithms.

\subsection{Visual Analytics for Neural Networks \& Predictions}

Many visual analytics tools have been developed to help deep learning experts analyze their models and predictions \cite{hohmanVisualAnalyticsDeep2019, garciaTaskandtechniqueCenteredSurvey2018, kahngActiVisVisualExploration2018a, liuBetterAnalysisDeep2017a, liuVisualizingHighDimensionalData2017, bilalConvolutionalNeuralNetworks2018}.
These tools support many different tasks.
For example, recent work such as Summit \cite{hohmanSummitScalingDeep2020} uses interactive visualization to summarize what features a CNN model has learned and how those features interact and attribute to model predictions.
LSTMVis\cite{strobeltLSTMVisToolVisual2018} makes long short-term memory (LSTM) networks more interpretable by visualizing the model's hidden states.
Similarly, GANVis\cite{wangGANVizVisualAnalytics2018} helps experts to interpret what a trained generative adversarial network (GAN) model has learned.
People also use visual analytics tools to diagnose and monitor the training process of deep learning models.
Two examples, DGMTracker\cite{liuAnalyzingTrainingProcesses2018} and DeepEyes\cite{pezzottiDeepEyesProgressiveVisual2018}, help developers better understand the training process of CNNs and GANs, respectively.
Also, visual analytics tools recently have been developed to help experts detect and interpret the vulnerability in their deep learning models\cite{liuAnalyzingNoiseRobustness2018a, dasMassifInteractiveInterpretation2020a}.
These existing analytics tools are designed to assist experts in analyzing their model and predictions, however, we focus on non-experts and learners, helping them more easily learn about deep learning concepts.

\section{Formative Research \& Design Challenges}
\label{sec:design-challenge}

Our goal is to build an interactive visual learning tool to help students gain understanding of key CNN concepts to design their own models.
To identify the learning challenges faced by the students, we conducted interviews with deep learning instructors and surveyed past students.

\begin{figure}[!b]
    \centering
    \includegraphics[width=0.95\columnwidth]{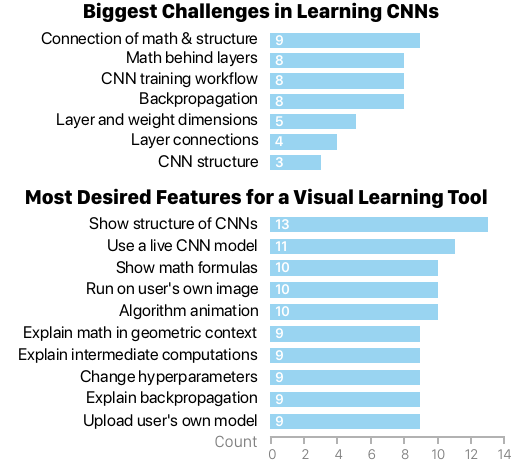}
    \caption{Survey results from 19 participants who have previously learned about CNNs.
    \textbf{Top:} Biggest challenges encountered during learning.
    \textbf{Bottom:} Desired features for an interactive visual learning tool for CNNs.}
    \label{fig:survey-result}
\end{figure}

\textbf{Instructor interviews.}
To inform our tool's design, we recruited 4 instructors (2 female, 2 male) who have taught CNNs in a large university.
We refer to them as T1-T4 throughout our discussion.
One instructor teaches computer vision, and the others teach deep learning.
We interviewed them one-on-one in a conference room (3/4) and via a video-conferencing software (1/4); each interview lasted around 30 minutes.
Through these semi-structured interviews, we learned that (1) instructors currently rely on simple illustrations with toy examples to explain CNN concepts, and an interactive tool like TensorFlow Playground with real image inputs would be highly appreciated; and (2) key challenges exist for instructors teaching and students learning about CNNs, which informed us to design a student survey.

\textbf{Student survey.}
After the interviews, we recruited students from a large university who have previously studied CNNs to fill out an online survey.
We received 43 responses, and 19 of them (4 female, 15 male) met the criteria.
Among these 19 participants, 10 were Ph.D. students, 3 were M.S. students, 5 were undergraduates, and 1 was a faculty member.
We asked participants what were ``the biggest challenges in studying CNNs'' and ``the most helpful features if there was a visualization tool for explaining CNNs to beginners''.
We provided pre-selected options based on the prior instructor interviews, but participants could write down their own response if it was not included in the options.
The aggregated results of this survey are shown in \autoref{fig:survey-result}.

Together with a literature review, we synthesized our findings from these two studies into the following five design challenges (C1-C5).

\begin{enumerate}[topsep=0mm, itemsep=0mm, parsep=1mm, leftmargin=6mm, label=\textbf{C\arabic*.}, ref=\textbf{C\arabic*}]
    \item \label{item:c1}
    \textbf{Intricate model structure.}
    CNN models consist of many layers, each having a different structure and underlying mathematical functions\cite{lecunDeepLearning2015}.
    Fewer past students listed CNN structure as their biggest challenge, but most of them believe a visual learning tool should explain the structure (\autoref{fig:survey-result}), as the complex construction of CNNs can be overwhelming, especially for beginners who just started learning.
    T2 said \quotes{It can be very hard for them [students with less knowledge of neural networks] to understand the structure of CNNs, you know, the connections between layers.}
    
    \item \label{item:c2}
    \textbf{Complex layer operations.}
    Different layers serve different purposes in CNNs\cite{guRecentAdvancesConvolutional2018}.
    For example, convolutional layers exploit the spatially local correlations in inputs---each convolutional neuron connects to only a small region of its input; whereas max pooling layers introduce regularization to prevent overfitting.
    T1 said, \quotes{The most challenging part is learning the math behind it [CNN model].}
    Many students also reported that CNN layer computations are the most challenging learning objective (\autoref{fig:survey-result}).
    To make CNNs perform better than other models in tasks like image classification, these models have complex and unique mathematical operations that many beginners may not have seen elsewhere.
    
    \item \label{item:c3}
    \textbf{Connection between model structure and layer operation.}
    Based on instructor interviews and the survey results from past students 
    (\autoref{fig:survey-result}), one of the cruxes to understand CNNs is understanding the interplay between low-level mathematical operations (\ref{item:c2}) and the high-level model structure (\ref{item:c1}).
    Smilkov et al., creators of the popular dense neural network learning tool Tensorflow Playground\cite{smilkovDirectManipulationVisualizationDeep2017}, also found this challenge key to learning about deep learning models: ``\textit{It's not trivial to translate the equations defining a deep network into a mental model of the underlying geometric transformations [change of feature representations].}''
    In other words, in addition to comprehending the mathematical formulas behind different layers, students are also required to understand how each operation works within the complex, layered model structure.

    \item \label{item:c4}
    \textbf{Effective algorithm visualization (AV).}
    The success of applying visualization to explain machine learning algorithms to beginners \cite{carneyTeachableMachineApproachable2020, smilkovDirectManipulationVisualizationDeep2017, kahngGANLabUnderstanding2019} suggests that an AV tool is a promising approach to help people more easily learn about CNNs.
    However, AV tools need to be carefully designed to be effective in helping learners gain an understanding of algorithms \cite{fouhRoleVisualizationComputer2012}.
    In particular, AV systems need to clearly explain the mapping between the algorithm and its visual encoding \cite{mayerAnimationsNeedNarrations1991}, and actively engage learners \cite{kehoeRethinkingEvaluationAlgorithm2001}.
    
    \item \label{item:c5}
    \textbf{Challenge in deploying interactive learning tools.}
    Most neural networks are written in deep learning frameworks, such as TensorFlow\cite{abadiTensorFlowSystemLargeScale2016} and PyTorch\cite{paszkePyTorchImperativeStyle2019a}.
    Although these libraries have made it much easier to create AI models, they require users to understand key concepts of deep learning in the first place \cite{stevensDeepLearningPyTorch2019}.
    Can we make understanding CNNs more accessible without installation and coding, so that everyone has the opportunity to learn and interact with deep learning models?

\end{enumerate}

\noindent
The above design challenges cover most of the desired features (\autoref{fig:survey-result}).
We assessed the feasibility to also support explaining backpropagation in the same tool, 
and we concluded that its effective explanation will 
necessitate designs that are hard to be unified (e.g., backpropagation Algorithm\cite{BackpropagationAlgorithm}) . 
Indeed, T1 commented that ``\textit{Deriving backpropagation is applying a series chain rules [...] It doesn't really make sense to visualize the gradients [in our tool]}.''
Supporting the training process would require client-side in-browser computation on many data examples, which incur both high amount of data download and slow convergence (\cite{karpathyConvNetJSMNISTDemo2016, kahngGANLabUnderstanding2019}). 
Therefore, as the first prototype, we decided for \tool{} to focus on explaining inference after a model has been trained.
We plan to support the explanation for
backpropagation and training process as future work (\autoref{sec:future}).
\section{Design Goals}
\label{sec:design-goal}

Based on the identified design challenges (\autoref{sec:design-challenge}), we distill the following key design goals (\ref{item:g1}–\ref{item:g5}) for \tool{}, an interactive visualization tool to help students more easily learn about CNNs.

\begin{enumerate}[topsep=0mm, itemsep=0mm, parsep=1mm, leftmargin=6mm, label=\textbf{G\arabic*.}, ref=\textbf{G\arabic*}]
    \item \label{item:g1}
    \textbf{Visual summary of CNN models and data flow.} Based on the survey results, showing the structure of CNNs is the most desired feature for a visual learning tool (\autoref{fig:survey-result}).
    Therefore, to give users an overview of the structure of CNNs, we aim to create a visual summary of a CNN model by visualizing all layer outputs and connections in one view.
    This could help users to visually track how input image data are transformed to final class predictions through a series of layer operations (\ref{item:c1}). (\autoref{sec:overview})
  
    \item \label{item:g2}
    \textbf{Interactive interface for mathematical formulas.} Since CNNs employ various complex mathematical functions to achieve high classification performance, it is important for users to understand each mathematical operation in detail (\ref{item:c2}).
    In response, we would like to design an interactive interface for each mathematical formula, enabling users to examine and better understand the inner-workings of layers. (\autoref{sec:detail-view})
    
    \item \label{item:g3}
    \textbf{Fluid transition between different levels of abstraction.} To help users connect low-level layer mathematical mechanisms to high-level model structure (\ref{item:c3}), we would like to design a focus + context display of different views, and provide smooth transitions between them.
    By easily navigating through different levels of CNN model abstraction, users can get a holistic picture of how CNN works. (\autoref{sec:view-transition})
    
    \item \label{item:g4}
    \textbf{Clear communication and engagement.} Our goal is to design and develop an interactive system that is easy to understand and engaging to use so that it can help people to more easily learn about CNNs (\ref{item:c4}).
    We aim to accompany our visualizations with explanations to help users to interpret the graphical representation of the CNN model (\autoref{sec:explanation}), and we wish to actively engage learners through visualization customizations. (\autoref{sec:customization})
    
    \item \label{item:g5}
    \textbf{Web-based implementation.} To develop an interactive visual learning tool that is accessible for users without installation and coding (\ref{item:c5}), we would like to use modern web browsers as the platform to explain the inner-workings of a CNN model, where users can access directly on their laptops or tablets.
    We also open-source our code to support future research and development of deep learning educational tools. (\autoref{sec:implementation})

\end{enumerate} 
\section{Visualization Interface of \tool{}}
\label{sec:interface}

\begin{figure}[!b]
    \centering
    \includegraphics[width=\columnwidth]{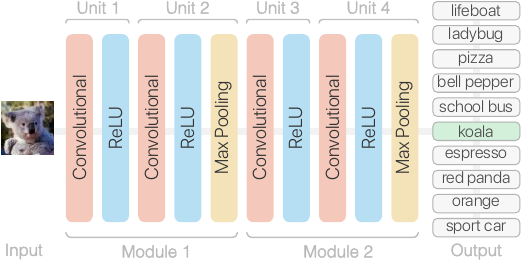}
    \caption{
    Illustration of \textit{Tiny VGG} model used in \tool{}:
    this model uses the same, but fewer, convolutional layers as the original VGGNet model \cite{simonyanVeryDeepConvolutional2015}.
    We trained it to classify 10 classes of images.
    }
    \label{fig:model}
\end{figure}

\begin{figure*}[!htb]
  \includegraphics[width=\textwidth]{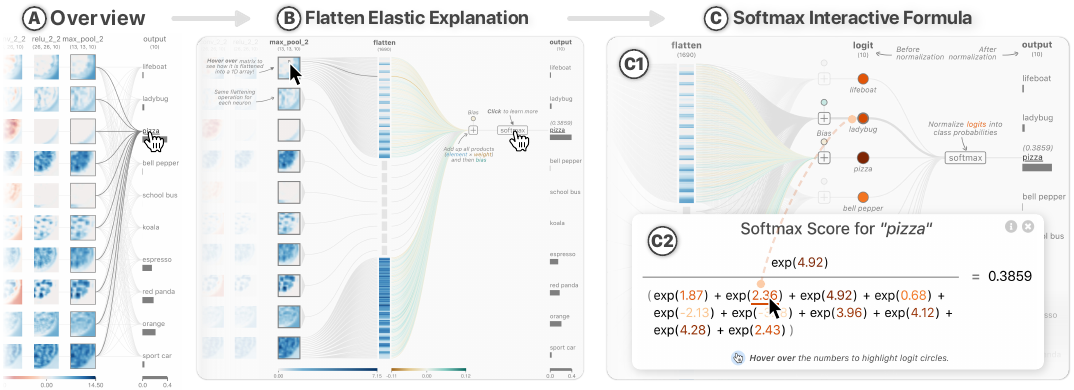}
  \caption{\tool{} helps users learn about the connection between the \layer{output} layer and its previous layer via three tightly integrated views.
  Users can smoothly transition between these views to gain a more holistic understanding of the \layer{output} layer's \protect\includegraphics[align=c, height=9pt]{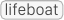} prediction computation.
  \textbf{(A)} The \textit{\overview{}} summarizes neurons and their connections. 
  \textbf{(B)} The \textit{\elasticFlatten{}} visualizes the often-overlooked flatten layer, helping users more easily understand how a high-dimensional \layer{max_pool_2} layer is connected to the 1-dimensional \layer{output} layer.
  \textbf{(C)} The \textit{\formulaSoft{}} further explains how the softmax function that precedes the \layer{output} layer normalizes the penultimate computation results (i.e., logits) into class probabilities through 
   linking the
  \textbf{(C1)} numbers from the formula to
  \textbf{(C2)} their visual representations within the model structure.
  }
  \label{fig:softmax-transition}
\end{figure*}

\tool{}'s interface is built on our prior prototype \cite{wangCNN101Interactive2020a}.
We visualize the forward propagation, i.e., transforming an input image into a class prediction, of a trained model (\autoref{fig:model}).
Users can explore a CNN at different levels of abstraction through the tightly integrated \textit{\overview{}} (\autoref{sec:overview}), \textit{\elastic{}} (\autoref{sec:intermediate-view}), and the \textit{\formula} (\autoref{sec:detail-view}).
Our tool allows users to smoothly transition between these views (\autoref{sec:view-transition}), provides text annotations and a tutorial article to help users interpret the visualizations (\autoref{sec:explanation}), and engages them to test hypotheses through visualization customizations (\autoref{sec:customization}).
The system is targeted towards beginners and describes all mathematical operations necessary for a CNN to classify an image.

\newcommand{\colorScaleFigure}[0]{
\setlength{\columnsep}{8pt}%
\setlength{\intextsep}{0pt}%
\begin{wrapfigure}{r}{0.15\textwidth}
    \vspace{2pt}
    \centering
    \includegraphics[width=0.15\textwidth]{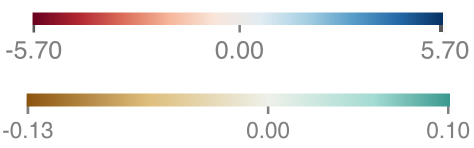}
\end{wrapfigure}
}

\colorScaleFigure{}
\textbf{Color scales} are used throughout the visualization to show the impact of weight, bias, and activation map values.
Consistently in the
interface, a \textbf{\textcolor{RdColorscale}{red}} to \textbf{\textcolor{BuColorscale}{blue}}
color scale is used to visualize neuron activation maps as heatmaps, and a \textbf{\textcolor{BrColorscale}{yellow}} to \textbf{\textcolor{BGColorscale}{green}} color scale represents weights and biases.
A persistent color scale legend is present across all views, so the user always has context for the displayed colors.
We chose these distinct, diverging color scales with white representing zero, so that a user can easily differentiate positive and negative values.
We group layers in the \textit{Tiny VGG} model, our CNN architecture, into four units and two modules (\autoref{fig:model}).
Each unit starts with one convolutional layer.
Both modules are identical and contain the same sequence of operations and hyperparameters.
To analyze neuron activations throughout the network with varying contexts, users can alter the range of the heatmap color scale (\autoref{sec:customization}).

\subsection{\overview{}}
\label{sec:overview}
The \textit{\overview{}} (\autoref{fig:crown}A, \autoref{fig:softmax-transition}A) is the opening view of \tool{}.
This view represents the high-level structure of a CNN: neurons grouped into layers with distinct, sequential operations.
It shows neuron activation maps for all layers represented as heatmaps with a diverging \textbf{\textcolor{RdColorscale}{red}} to \textbf{\textcolor{BuColorscale}{blue}} color scale.
Neurons in consecutive layers are connected with edges, which connect each neuron to its inputs; to see these edges, users simply can hover over any activation map.
In the model, neurons in convolutional layers and the \layer{output} layer are fully connected to the previous layer, while all other neurons are only connected to one neuron in the previous layer.

\subsection{Elastic Explanation View}
\label{sec:intermediate-view}
The \textit{\elastic{}s} visualize the computations that leads to an intermediate result without overwhelming users with low-level mathematical operations.
\tool{} enters two elastic views after a user clicks a convolutional or an output neuron from the \textit{\overview{}}.
After the transition, far-away heatmaps and edges fade out to help users focus on the selected layers while providing CNN structural context in the background (\autoref{fig:crown}A).

\textbf{Explaining the Convolutional Layer} (\autoref{fig:crown}B).
The \textit{\elasticConv{}} applies a convolution on each input node of the selected neuron, visualized by a kernel sliding across the input neurons, which yields an intermediate result for each input neuron.
This sliding kernel forms the output heatmap during the animation, which imitates the internal process during a convolution operation.
While the sliding kernel animation is in progress, the edges in this view are represented as flowing-dashed lines; upon the animations completion, the edges transition to solid lines.

\textbf{Explaining the Flatten Layer} (\autoref{fig:softmax-transition}B).
The \textit{\elasticFlatten{}} visualizes the operation of transforming an n-dimensional tensor into a 1-dimensional tensor by traversing pixels in row-major order.
This flattening operation is often necessary in a CNN prior to classification so that the fully-connected \layer{output} layer can make classification decisions.
The view represents each neuron in the \layer{flatten} layer as a short line whose color is the same as its source pixel in the previous layer.
Then, edges connect these neurons with their source components and intermediate results.
These edges are colored based on the model's weight value.
Users can hover over any component of this connection to highlight the associated edges as well as the \layer{flatten} layer's neuron and the pixel value from the previous layer.

\begin{figure}[!tb]
    \centering
    \includegraphics[width=\columnwidth]{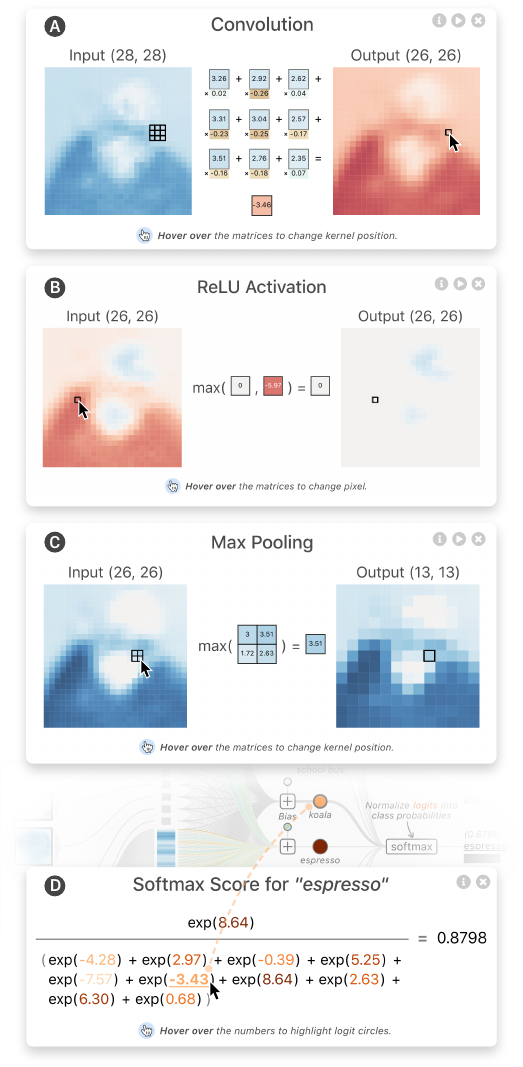}
    \caption{The \textit{\formula{}s} explain the underlying mathematical operations of a CNN.
    \textbf{(A)} shows the element-wise dot-product occurring in a convolutional neuron,
    \textbf{(B)} visualizes the activation function ReLU, and 
    \textbf{(C)} illustrates how max pooling works.
    Users can hover over heatmaps to display an operation's input-to-output mapping.
    \textbf{(D)} interactively explains the softmax function, helping users connect numbers from the formula to their visual representations.
    Users can click the info button \protect\includegraphics[align=c, height=8pt]{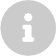} to scroll to the corresponding section in the tutorial article, and the play button \protect\includegraphics[align=c, height=8pt]{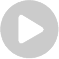} to start the window sliding animation in (A)-(C).
    }
    \label{fig:detail-view}
\end{figure}

\subsection{\formula{}}
\label{sec:detail-view}
The \textit{\formula{}} consists of four variations designed for convolutional layers, ReLU activation layers, pooling layers, and the softmax activation function.  
After users have built up a mental model of the CNN model structure from the previous \textit{\overview{}} and \textit{Elastic Explanation Views}, these four views demonstrate the detailed mathematics occurring in each layer.

\textbf{Explaining Convolution, ReLU Activation, and Pooling} (\autoref{fig:detail-view}A, B, C))
Each view animates the window-sliding operation on the input matrix and output matrix over an interval, so that the user can understand how each element in the input is connected to the output, and vice versa.
In addition, the user can interact with the these matrices by hovering over the heatmaps to control the position of the sliding window.
For example, in the \textit{\formulaConv{}} (\autoref{sec:detail-view}A), as the user controls the window (kernel) position in either the input or the output matrix, this view visualizes the dot-product formula with input numbers and kernel weights directly extracted from the current kernel.
This synchronization between the input, the output and the mathematical function enables the user to better understand how the kernel convolves a matrix in convolutional layers.

\textbf{Explaining the Softmax Activation} (\autoref{fig:detail-view}D).
This view outlines the operations necessary to calculate the classification score.
It is accessible from the \textit{\elasticFlatten{}} to explain how the results (logits) from the previous view lead to the final classification.
The view consists of logit values encoded as circles and colored with a \textbf{\textcolor{LOrangesColorscale}{light orange}} to \textbf{\textcolor{ROrangesColorscale}{dark orange}} color scale, which provides users with a visual cue of the importance of every class.
This view also includes a corresponding equation, which explains how the classification score is computed.
When users enter this view, pairs of each logit circle and its corresponding value in the equation appear sequentially with animations.
As a user hovers over a logit circle, its value will be highlighted in the equation along with the logit circle itself, so the user can understand how each logit contributes to the softmax function.
Hovering over numbers in the equation will also highlight the appropriate logit circles.
Interacting with logit circles and the mathematical equation in combination allows a user to discern the impact that every logit has on the classification score in the \layer{output} layer.

\subsection{Transitions Between Views}
\label{sec:view-transition}
The \textit{\overview{}} is the starting state of \tool{} and shows the model architecture.
From this high-level view, the user can begin inspecting layers, connectivity, classifications, and tracing activations of neurons through the model.
When a user is interested in more detail, they can click on neuron activation maps in the visualization.
Neurons in a layer that have simple one-to-one connections to a neuron in the previous layer do not require an auxiliary \textit{\elastic{}}, so upon clicking one of these neurons, a user will be able to enter the \textit{\formula{}} to understand the low-level operation that a tensor undergoes at that layer.
If a neuron has more complex connectivity, then the user will enter an \textit{\elastic{}} first.
In this view, \tool{} uses visualizations and annotations before displaying mathematics.
Through further interaction, a user can hover and click on parts of the \textit{\elastic{}} to uncover the mathematical operations as well as examine the values of weights and biases.
The low-level \textit{\formula{}s} are only shown after transitioning from the previous two views, so that users can learn about the underlying mathemtical operations after hainvg a mental model of the complex and layered CNN model structure.

\subsection{Visualizations with Explanations}
\label{sec:explanation}

\tool{} is accompanied by an interactive tutorial article beneath the interface that explains CNN layer functions, hyperparameters, and outlines \tool{}'s interactive features.
Learners can read freely, or jump to specific sections by clicking layer names or the info buttons (\autoref{fig:detail-view}) from the main visualization.
The article provides beginner users detailed information regarding CNNs that can supplement their exploration of the visualization.

\setlength{\columnsep}{8pt}%
\setlength{\intextsep}{0pt}%
\begin{wrapfigure}{R}{0.13\textwidth}
    \centering
    \includegraphics[width=0.13\textwidth]{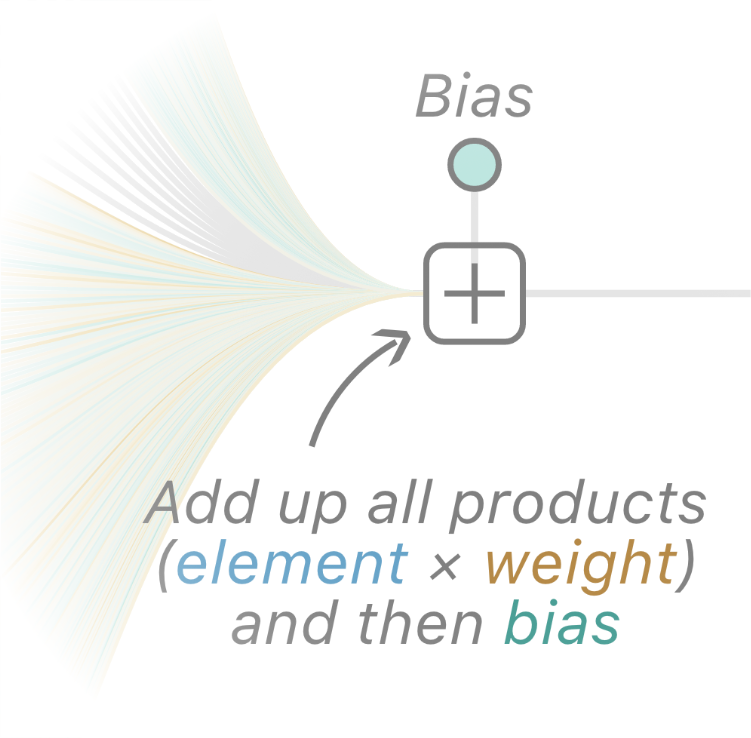}
\end{wrapfigure}
Additionally, text annotations are placed throughout the visualization (e.g., explaining the flatten layer operation in the right image), which further guide users and explain concepts that are not easily discernible from the visualization alone.
These annotations help users map the underlying algorithm to its visual encoding.

\subsection{Customizable Visualizations}
\label{sec:customization}
\begin{figure}[tb]
    \centering
    \includegraphics[width=\columnwidth]{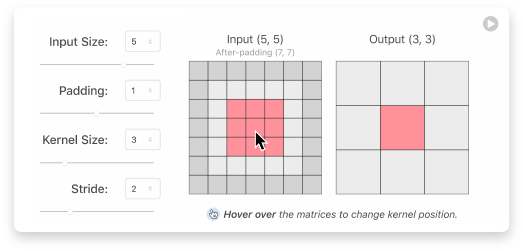}
    \caption{The \textit{\widgetParam}, a component of the accompanying interactive article, allows users to adjust hyperparameters and observe in real time how the kernel's sliding pattern changes in convolutional layers.}
    \label{fig:hyperparameter-visualization}
\end{figure}

The \textit{Control Panel} located across the top of the visualization (\autoref{fig:crown}) allows the user to alter the CNN input image and edit the overall representation of the network. 
The \textit{\widgetParam{}} (\autoref{fig:hyperparameter-visualization}) enables the user to experiment with differnt convolution hyperparameters.

\setlength{\columnsep}{8pt}%
\setlength{\intextsep}{0pt}%
\begin{wrapfigure}{R}{0.16\textwidth}
    \vspace{0pt}
    \centering
    \includegraphics[width=0.16\textwidth]{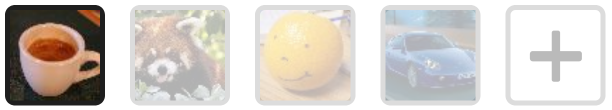}
\end{wrapfigure}
\textbf{Change input image.}
Users can choose between (1) preloaded input images for each output class, or (2) upload their own custom image.
Preloaded images allow a user to easily access data from the classes the model was originally trained on.
User can also freely upload any image for classification into the ten classes the network was trained on.
\tool{} resizes a user's image while preserving the aspect ratio to fit one dimension of the model input size, and then crop the central region if the other dimensions does not match.
The fourth of six AV tool engagement levels is allowing users to change the AV tool's input\cite{napsExploringRoleVisualization2002}.
Supporting custom image upload engages users, by allowing them to analyze the network's classification decisions and interactively testing their own hypotheses on diverse image inputs.

\setlength{\columnsep}{8pt}%
\setlength{\intextsep}{0pt}%
\begin{wrapfigure}{R}{0.1\textwidth}
    \centering
    \includegraphics[width=0.1\textwidth]{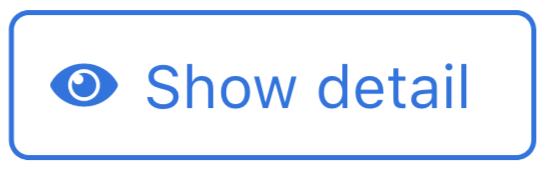}
\end{wrapfigure}
\textbf{Show network details.}
A user can toggle the ``Show detail'' button, which displays additional network specifications in the \textit{\overview{}}.
When toggled on, the \textit{\overview{}} will reveal layer dimensions and show color scale legends.
Additionally, a user can vary the activation map color scale range.
The CNN architecture presented by \tool{} is grouped into four units and two modules (\autoref{fig:model}).
By modifying the drop-down menu in the \textit{Control Panel}, a user can adjust the color scale range used by the network to investigate activations with different groupings.

\textbf{Explore hyperparameter impact.}
The tutorial article (\autoref{sec:explanation}) includes an interactive \textit{\widgetParam{}} that allows users to experiment with convolutional hyperparameters (\autoref{fig:hyperparameter-visualization}).
Users can adjust the input and hyperparameters of the stand-alone visualization to test how different hyperparameters change the sliding convolutional kernel and the output's dimensions.
This interactive element emphasizes learning through experimentation by supplementing knowledge gained from reading the article and using the main visualization.

\subsection{Web-based, Open-sourced Implementation}
\label{sec:implementation}
\tool{} is a web-based, open-sourced visualization tool to teach students the foundations of CNNs.
A new user only needs a modern web-broswer to access our tool, no installation required.
Additionally, other datasets and linear models can be quickly applied to our visualization system due to our robust implementation.

\textbf{Model Training.}  The CNN architecture, Tiny VGG (\autoref{fig:model}), presented by \tool{} for image classification is inspired by both the popular deep learning architecture, VGGNet \cite{simonyanVeryDeepConvolutional2015}, and Stanford's CS231n course notes \cite{karpathyCS231nConvolutionalNeural2016}.
It is trained on the Tiny ImageNet dataset \cite{TinyImageNetVisual2015}.
The training dataset consists of 200 image classes and contains 100,000 64$\times$64 RGB images, while the validation dataset contains 10,000 images across the 200 image classes.
The model is trained using \textit{TensorFlow} \cite{abadiTensorFlowSystemLargeScale2016} on 10 handpicked, everyday classes: \protect\includegraphics[align=c, height=9pt]{figures/symbol-lifeboat.pdf},
\protect\includegraphics[align=c, height=9pt]{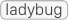},
\protect\includegraphics[align=c, height=9pt]{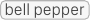},
\protect\includegraphics[align=c, height=9pt]{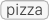},
\protect\includegraphics[align=c, height=9pt]{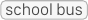},
\protect\includegraphics[align=c, height=9pt]{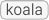},
\protect\includegraphics[align=c, height=9pt]{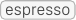},
\protect\includegraphics[align=c, height=9pt]{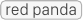},
\protect\includegraphics[align=c, height=9pt]{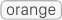},
and \protect\includegraphics[align=c, height=9pt]{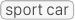}.
During the training process, the batch size and learning rate are fine-tuned using a 5-fold-cross-validation scheme.
This simple model achieves a 70.8\% top-1 accuracy on the validation dataset.  

\textbf{Front-end Visualization.}
\tool{} loads the pre-trained Tiny VGG model and computes forward propagation results in real time in a user's web browser using  \textit{TensorFlow.js} \cite{smilkovTensorFlowJsMachine2019}.
These results are visualized using \textit{D3.js} \cite{bostockDataDrivenDocuments2011} throughout the multiple interactive views.
\section{Usage Scenarios}

\subsection{Beginner Learning Layer Connectivity}
\label{sec:scenarioResearcher}
Janis is a virology researcher using CNNs in a current project.
Through an online deep learning course she has a general understanding of the goals of applying CNNs, and some basic knowledge of different types of CNN layers, but she needs help filling in some gaps in knowledge.
Interested in learning how a 3-dimensional input (RGB image) leads to a 1-dimensional output (vector of class probabilities) in a CNN, Janis begins exploring the architecture from the \textit{\overview{}} (\autoref{fig:softmax-transition}A).

After clicking the ``Show detail'' button, Janis notices that the \layer{output} layer is a 1-dimensional tensor of size 10, while \layer{max_pool_2}, the previous layer, is a 3-dimensional (13$\times$13$\times$10) tensor.
Confused, she hovers over a neuron in the \layer{output} layer to inspect connections between the final two layers of the architecture: the \layer{max_pool_2} layer has 10 neurons; the \layer{output} layer has 10 neurons each representing a class label, and the \layer{output} layer is fully-connected to the \layer{max_pool_2} layer.
She clicks that \layer{output} neuron, which transitions the \textit{\overview{}} (\autoref{fig:softmax-transition}A) to the \textit{\elasticFlatten} (\autoref{fig:softmax-transition}B).
She notices that edges between these two layers intersect a 1-dimensional flatten layer and pass through a softmax function.
By hovering over pixels from the activation map, Janis understands how the 2-dimensional matrix is ``unwrapped'' to yield a portion of the 1-dimensional \layer{flatten} layer.
As she continues to follow the edge after the \layer{flatten} layer, she clicks the softmax button which leads her to the \textit{\formulaSoft{}} (\autoref{fig:softmax-transition}C).
She learns how the outputs of the \layer{flatten} layer are normalized by observing the equation linked with logits through animations.
Janis recognizes that her previous coursework has not taught these ``hidden'' operations prior to the \layer{output} layer, which flatten and normalize the output of the \layer{max_pool_2} layer.
Instead of searching through lecture videos and textbooks, \tool{} enables Janis to learn these often-overlooked operations through a hierarchy of interactive views in a stand-alone website.
She now feels more equipped to apply CNNs to her virology research.

\subsection{Teaching Through Interactive Experimentation}
\label{sec:scenarioTeaching}
A university professor, Damian, is currently teaching a computer vision class which covers CNNs.
Damian begins his lecture with standard slides.
After describing the theory of convolutions, he opens \tool{} to demonstrate the convolution operation working inside a full CNN for image classification.
With \tool{} projected to the class, Damian transitions from the \textit{\overview{}} (\autoref{fig:crown}A) to the \textit{\elasticConv{}} (\autoref{fig:crown}B).
Damian encourages the class to interpret the sliding window animation (\autoref{fig:intro-figure}B) as it generates several intermediate results.
He then asks the class to predict kernel weights in a specific neuron.
To test student's hypotheses, Damian enters the \textit{\formulaConv{}} (\autoref{fig:crown}C), to display the convolution operation with the true kernel weights.
In this view, he can hover over the input and output matrices to answer questions from the class, and display computations behind the operation.

Recalled from theory, a student asks a question regarding the impact of altering the stride hyperparameter on the animated sliding window in convolutional layers.
To illustrate the impact of alternative hyperparameters, Damian scrolls down to the ``Convolutional Layer'' section of the complementary article, where he experiments by adjusting stride and other hyperparameters with the \textit{\widgetParam{}} (\autoref{fig:hyperparameter-visualization}) in front of the class.
\tool{} is the first software that allows Damian to explain convolutional operations and hyperparameters with real image inputs, and quickly answer students' questions in class.
Previously, Damian had to draw illustrations with simple matrix inputs on slides or a chalkboard.
Finally, to reinforce the concepts and encourage 
individual 
experimentation, Damian provides the class with a URL to 
the web-based \tool{} for students to return to in the future.
\section{Observational Study}
\label{sec:user-study}

We conducted an observational study to investigate how \tool{}'s target users (e.g., aspiring deep learning students) would use this tool to learn about CNNs, and also to test the tool's usability.

\subsection{Participants}

\tool{} is designed for deep learning beginners who are interested in learning CNNs.
In this study, we aimed to recruit participants who aspire to learn about CNNs and have some knowledge of basic machine learning concepts (e.g., knowing what an image classifier is).
We recruited 16 student participants from a large university (4 female, 12 male) through internal mailing lists (e.g., machine learning and  computer science Ph.D., M.S., and undergraduate students).
Seven participants were Ph.D. students, seven were M.S. students, and the other two were undergraduates.
All participants were interested in learning CNNs, and none of them had known \tool{} before.
Participants self-reported their level of knowledge on non-neural network machine learning techniques, with an average score of 3.26 on a scale of 0 to 5 (0 being ``no knowledge'' and 5 being ``expert''); and an average score of 2.06 on CNNs (on the same scale).
No participant self-reported a score of 5 for their knowledge on CNNs, and one participant had a score of 0.
To help better organize our discussion, we refer to participants with CNN knowledge score of 0, 1 or 2 as B1-B11, where ``B'' stands for ``Beginner''; 
and those with score of 3 or 4 as K1-K5, where ``K'' stands for ``Knowledgeable.''

\subsection{Procedure}

We conducted this study with participants one-on-one via video-conferencing software.
With the permission of all participants, we recorded the participants' audio and computer screen for subsequent analysis.
After participants signed consent forms, we provided them a 5-minute overview of CNNs, followed by a 3-minute tutorial of \tool{}.
Participants then freely explored our tool in their computer's web browser.
We also provided a feature checklist, which outlined the main features of our tool and encouraged participants to try as many features as they could.
During the study, participants were asked to think aloud and share their computer screen with us; they were encouraged to ask questions when necessary.
Each session ended with a usability questionnaire coupled with an exit interview that asked participants about their process of using \tool{}, and if this tool could be helpful for them.
Each study lasted around 50 minutes, and we compensated each participant with a \$10 Amazon Gift card.

\subsection{Results and Design Lessons}
\label{sec:study-result}

\begin{figure}[tb]
    \centering
    \includegraphics[width=\columnwidth]{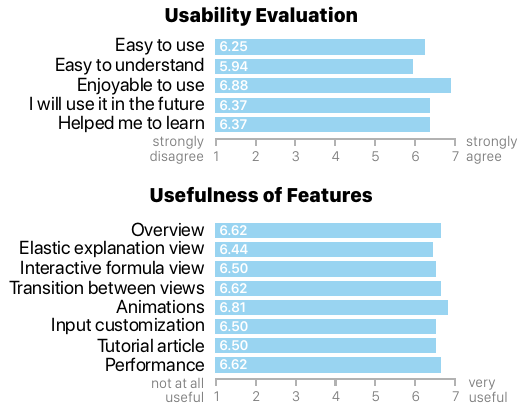}
    \caption{Average ratings from 16 participants regarding the usability and usefulness of \tool{}.
    \textbf{Top:} Participants thought \tool{} was easy to use, enjoyable, and helped them learn about CNNs.
    \textbf{Bottom:} All features, especially animations, were rated favorably.
    }
    \label{fig:study-result}
\end{figure}

The exit questionnaire included a series of 7-point Likert-scale questions about the utility and usefulness of different views in \tool{} (\autoref{fig:study-result}).
All average Likert rating were above 6 except the rating of ``easy to understand''.
From the high ratings and our observations, participants found our tool easy to use and understand, retained a high engagement level during their session, and eventually gained a better understanding of CNN concepts.
Our observations also reflect key findings in previous AV research~\cite{fouhRoleVisualizationComputer2012, kehoeRethinkingEvaluationAlgorithm2001}.
This section describes design lessons and limitations of our tool distilled from this study.

\subsubsection{Transitions between different views}
\textbf{Transitions help users link CNN operations and structures.}
Several participants (9/16) commented that they liked how our tool transitions between high-level CNN structure views and low-level mathematical explanations.
It helps them better understand the interplay between layer computations and the overall CNN data transformation---one of the key challenges for understanding CNN concepts, as we identified from our instructor interviews and our student survey.
For example, initially K4 was confused to see the \textit{\elasticConv{}}, but after reading the annotation text, he remarked, \quotes{Oh, I understand what an intermediate layer is now---you run the convolution on the image, then you add all those results to get this.}
After exploring the \textit{\formulaConv{}}, he immediately noted, \quotes{Every single aspect of the convolution layer is shown here. [This] is super helpful.}
Similarly, B5 commented, \quotes{Good to see the big picture at once and the transition to different views [...] I like that I can hide details of a unit in a compact way and expand it when [needed].}

\tool{} employs the fisheye view technique for presenting the \textit{\elastic{}s} (\autoref{fig:crown}B, \autoref{fig:softmax-transition}B): after transitioning from the \textit{\overview{}} to a specific layer, neighboring layers are still shown while further layers (lower degree-of-interest) have lower opacity.
Participants found this transition design helpful for them to learn layer-specific details while having CNN structural context in the background.
For instance, K5 said \quotes{I can focus on the current layer but still know the same operation goes on for other layers.}
Our observations from this study suggest that our fluid transition design between different level of abstraction can help users to better connect unfamiliar layer mechanisms to the complex model structure.

\subsubsection{Animations for enjoyable learning experience}
Another favorite feature of \tool{} that participants mentioned was the use of animations, which received the highest rating in the exit questionnaire (\autoref{fig:study-result}).
In our tool, animations serve two purposes: to assimilate the relationship between different visual components and to help illustrate the model's underlying operations.

\textbf{Transition animations help navigating.}
Layer movement is animated during view transitions.
We noticed it helped participants to be aware of different views, and all participants navigated through the views naturally.
In addition to assisting with understanding the relationship between distinct views, animation also helped them discover the linking between different visualization elements.
For example, B8 quickly found that the logit circle is linked to its corresponding value in the formula, when she saw the circle-number pair appear one-by-one with animation in the \textit{\formulaSoft{}} (\autoref{fig:softmax-transition}C).

\textbf{Algorithm animations contribute to understanding.}
Animations that simulate the model's inner-workings helped participants learn underlying operations by validating their hypotheses. 
In the \textit{\elasticConv{}} (\autoref{fig:intro-figure}B), we animate a small rectangle sliding through one matrix to mimic the CNN's internal sliding window.
We noticed many participants had their attention drawn to this animation when they first transitioned into the \textit{\elasticConv{}}.
However, they did not report that they understood the convolution operation until interacting with other features, such as reading the annotation text or transitioning to the \textit{\formulaConv{}} (\autoref{fig:intro-figure}C).
Some participants went back to watch the animation multiple times and commented that it made sense, for example, K5 said \quotes{Very helpful to see how the image builds as the window slides through,} but others, such as B9 remarked, \quotes{It is not easy to understand [convolution] using only animation.}
Therefore, we hypothesize that this animation can indirectly help users to learn about the convolution algorithm by validating their newly formed mental models of how specific operation behave.
To test this hypothesis, a rigorous controlled experiment would be needed.
Related research work on the effect of animation in computer science education also found that algorithm animation does not automatically improve learning, but it may lead learners to make predictions of the algorithm behavior which in turn helps learning \cite{byrneEvaluatingAnimationsStudent1999}.

\textbf{Animations improve learning engagement and enjoyment.}
We found animations helped to increase participants' engagement level (e.g., spending more time and effort) and made \tool{} more enjoyable to use.
In the study, many participants repeatedly played and viewed different animations.
For example, K2 replayed the window sliding animation multiple times: \quotes{The is very well-animated [...] I always love smooth animations.}
B7 also attributed animations to his enjoyable experience with our tool: \quotes{[The tool is] enjoyable to use [...] I especially like the lovely animation.} 

\subsubsection{Engaging learning through visualization customization}
\tool{} allows users to modify the visualization.
For example, users can change the input image or upload their own image for classification; \tool{} visualizes the new prediction with the new activation maps in every layer.
Similarly, users can interactively explore how hyperparameters affect the convolution operation (\autoref{fig:hyperparameter-visualization}).

\setlength{\columnsep}{8pt}%
\setlength{\intextsep}{2pt}%
\begin{wrapfigure}{R}{0.08\textwidth}
    \centering
    \includegraphics[width=0.08\textwidth]{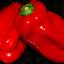}
\end{wrapfigure}

\textbf{Customization enables hypothesis testing.}
Many participants used visualization customization to test their predictions of model behaviors.
For example, through inspecting the input layer in the \textit{\overview{}}, B4 learned that the input layer comprised multiple different image channels (e.g., red, green, and blue).
He changed the input image to a red bell pepper from Tiny Imagenet (shown on the right) and expected to see high values in the input red channel: \quotes{If I click the red image, I would see...}
After the updated visualization showed what he predicted, he said \quotes{Right, it makes sense.}
We found the \textit{\widgetParam{}} also allowed participants to test their hypotheses.
While reading the description of convolution hyperparameters in the tutorial article, K3 noted \quotes{Wait, then sometimes they won't work}.
He then modified the hyperparatmeters in the \textit{\widgetParam{}} and noticed some combinations indeed did not yield a valid operation output: \quotes{It won't be able to slide, because the stride and kernel size don't fit the matrix}.

\setlength{\columnsep}{8pt}%
\setlength{\intextsep}{2pt}%
\begin{wrapfigure}{R}{0.08\textwidth}
    \centering
    \includegraphics[width=0.08\textwidth]{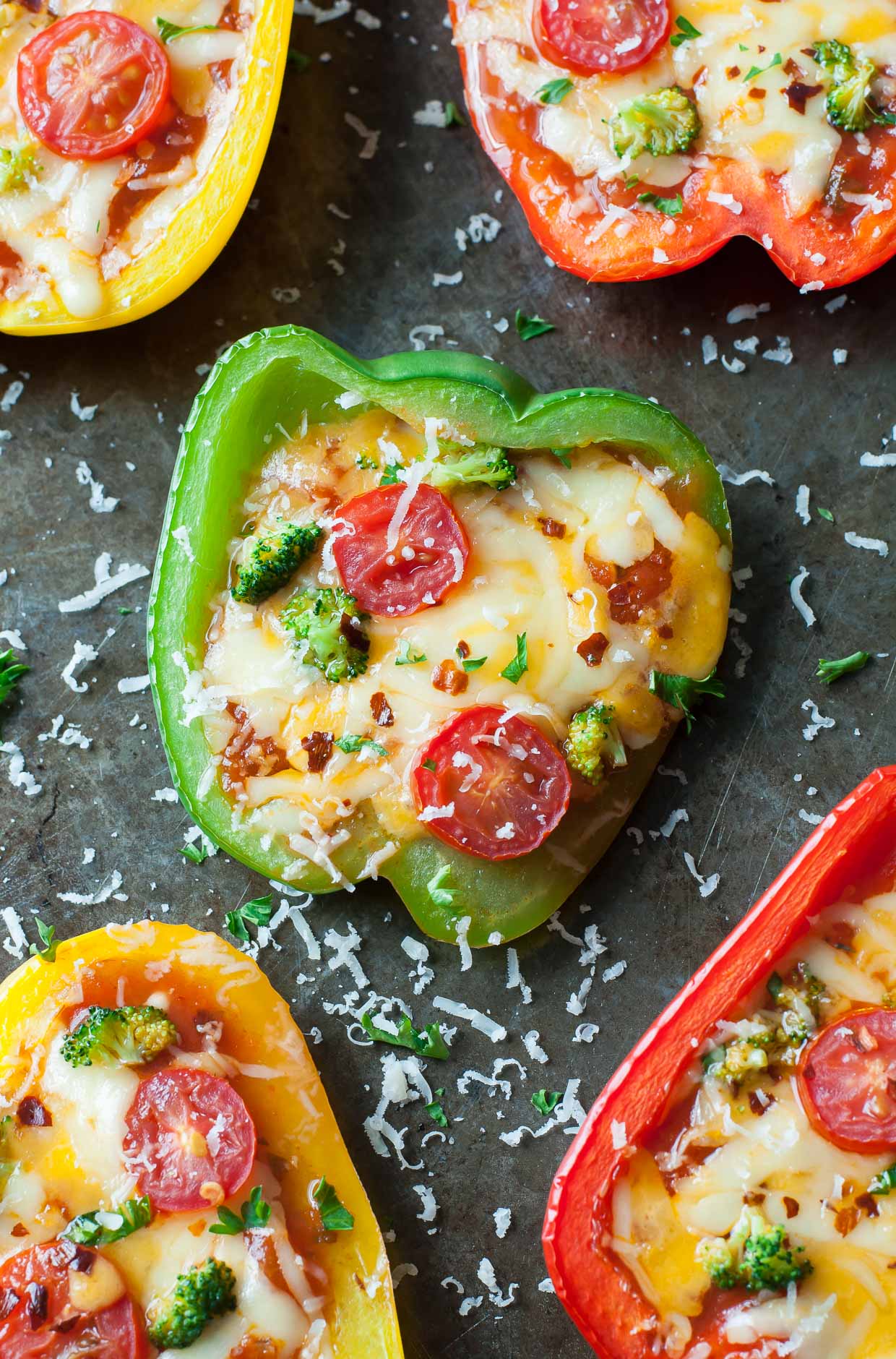}
\end{wrapfigure}

\textbf{Customization facilitates engagement.} Participants were intrigued to modify the visualization, and their engagement sparked further interest in learning CNNs.
In the study, B6 spent a large amount of time on testing the CNN's behavior on edge cases by finding ``difficult'' images online.
He searched with keywords ``koala'', ``koala in a car'', ``bell pepper pizza'', and eventually found a bell pepper pizza photo (shown on the right\footnote{Photo by Jennifer Laughlin, used with permission.}).
Our CNN model predicted the image as \protect\includegraphics[align=c, height=9pt]{figures/symbol-pepper.pdf} with a probability of $0.71$ and \protect\includegraphics[align=c, height=9pt]{figures/symbol-ladybug.pdf} with a probability of $0.2$.
He commented, \quotes{The model is not robust [...] oh, the ladybug ['s high softmax score] might come from the red dot.}
Another participant B5 uploaded his own photo as a new input image for the CNN model.
After seeing his picture being classified as \protect\includegraphics[align=c, height=9pt]{figures/symbol-espresso.pdf}, B5 started to use our tool to explore the reason of such classification by tracking back activation maps.
He also asked how do experts interpret CNNs and said he would be interested in learning more about deep learning interpretability.
This observation reflects previous findings that customizable visualization makes learning more engaging \cite{fouhRoleVisualizationComputer2012,napsExploringRoleVisualization2002}.

\subsubsection{Limitations}

While we found \tool{} provided participants with an engaging and enjoyable learning experience and helped them to more easily learn about CNNs, we also noticed some potential improvements to our current system design from this study.

\textbf{Beginners need more guidance.}
We found that participants with less knowledge of CNNs needed more instructions to begin using \tool{}.
Some participants reported that the visual representation of the CNN and animation initially were not easy to understand, but the tutorial article and text annotations greatly helped them to interpret the visualizations.
B8 skimmed through the tutorial article before interacting with the main visualization.
She said, \quotes{After going through the article, I think I will be able to use the tool better [...] I think the article is good, for beginner users especially.}
B2 appreciated the ability to jump to a certain section in the article by clicking the layer name in the visualization, and he suggested us to \quotes{include a step-by-step tutorial for first time users [...] There was too much information, and I didn't know where to click at the beginning}.
Therefore, we believe adding more text annotation and having a step-by-step tutorial mode could help
beginners better understand the relations between CNN operations and their visual representations.

\textbf{Limited explanation of \textit{why} CNN works.}
Some participants, especially those less experienced with CNNs, were interested in learning \textit{why} the CNN architecture works in addition to learning \textit{how} a CNN model makes predictions.
For example, B7 asked \quotes{Why do we need ReLU?} when he was learning the formula of the ReLU function.
B5 understood what a Max Pooling layer's operation does but was unclear why it contributes to CNN's performance: \quotes{It is counter-intuitive that Max Pooling reduces the [representation] size but makes the model better.}
Similarly, B6 commented on the Max Pooling layer: \quotes{Why not take the minimum value? [...] I know how to compute them [layers], but I don't know why we compute them.}
Even though it is still an open question why CNNs work so well for various applications \cite{guRecentAdvancesConvolutional2018, zhangUnderstandingDeepLearning2017a}, there are some commonly accepted ``intuitions'' of how different layers help this model class succeed.
We briefly explain them in the tutorial article: for example, ReLU function is used to introduce non-linearty in the model.
However, we believe it is worth designing visualizations that help users to learn about these concepts.
For example, allowing users to change the ReLU activation function to a linear function, and then visualizing the new model predictions may help users gain understanding of \textit{why} non-linear activation functions are needed in CNNs. 
\section{Discussion and Future Work}
\label{sec:future}

\indent\indent \textbf{Explaining training process and backpropagation.} \tool{} helps users to learn how a pre-trained CNN model transforms the input image data into a class prediction.
As we identified from two preliminary studies and an observational study, students are also interested in learning about the training process and backpropagation of CNNs.
We plan to work with instructors and students to design and develop new visualizations to help beginners gain understanding of the training process and backpropagation in detail.

\textbf{Generalizing to other layer types and neural network models.}
Our observational study demonstrated that \tool{} helps users more easily understand low-level layer operations, high-level model structure, and their connections.
We can adapt the \textit{\formula{}s} to explain other layer types (e.g., Leaky ReLU \cite{Maas13rectifiernonlinearities}) or a combination of layers (e.g. Residual Block \cite{heDeepResidualLearning2015}).
Similarly, the transition between different levels of abstraction can be generalized to other neural networks, such as long short-term memory networks \cite{hochreiterLongShortTermMemory1997} and Transformer models \cite{vaswaniAttentionAllYou2017} that require learners to understand the intricate layer operations in the context of a complex network structure.

\textbf{Integrating algorithm visualization best practices.} Existing work has studied how to design effective visualizations to help students learn algorithms.
\tool{} applies two key design principles from AV---visualizations with explanations and customizable visualizations (\ref{item:g4}).
However, there are many other AV design practices that future researchers can integrate in educational deep learning tools, such as giving interactive ``pop quizzes'' during the visualization process \cite{napsJHAVEEnvironmentActively2000} and encouraging users to build their own visualizations \cite{staskoUsingStudentbuiltAlgorithm1997}.

\textbf{Quantitative evaluation of educational effectiveness.}
We conducted a qualitative observational study to evaluate the usefulness and usability of \tool{}.
Further quantitative user studies would help us investigate how visualization tools help users gain understanding of deep learning concepts.
We will draw inspiration from recent research \cite{kahngHowDoesVisualization2019, conlenCaptureAnalysisActive2019} to assess users' engagement level and content understanding through analysis of interaction logs. 
\section{Conclusion}

As deep learning is increasingly used throughout our everyday life, it is important to help learners take the first step toward understanding this promising yet complex technology.
In this work, we present \tool{}, an interactive visualization system designed for non-experts to more easily learn about CNNs.
Our tool runs in modern web browsers and is open-sourced, broadening the public's education access to modern AI techniques.
We discussed design lessons learned from our iterative design process and an observational user study.
We hope our work will inspire further research and development of visualization tools that help democratize and lower the barrier to understanding and appropriately applying AI technologies. 
\acknowledgments{
We thank Anmol Chhabria, Kaan Sancak, Kantwon Rogers, and the Georgia Tech Visualization Lab for their support and constructive feedback.
This work was supported in part by NSF grants IIS-1563816, CNS-1704701, NASA NSTRF, DARPA GARD, gifts from Intel, NVIDIA, Google, Amazon.
Use, duplication, or disclosure is subject to the restrictions as stated in Agreement number HR00112030001 between the Government and the Performer.
} 
\bibliographystyle{abbrv-doi}

\bibliography{cnn-explainer}
\end{document}